\documentclass[12pt,preprint]{aastex}

\usepackage{xcolor}

\shorttitle{}
\shortauthors{Nesvorn\'y}

\begin{document}
\baselineskip 19.pt

\title{Eccentric Early Migration of Neptune}

\author{David Nesvorn\'y}
\affil{Department of Space Studies, Southwest Research Institute,\\
1050 Walnut St., Suite 300, Boulder, CO, 80302, USA}

\begin{abstract} 

The dynamical structure of the Kuiper belt can be used as a clue to the formation and evolution of the Solar 
System, planetary systems in general, and Neptune's early orbital history in particular. The problem is best 
addressed by forward modeling where different initial conditions and Neptune's orbital evolutions are tested, 
and the model predictions are compared to orbits of known Kuiper belt objects (KBOs). It has previously been 
established that Neptune radially migrated, by gravitationally interacting with an outer disk of planetesimals, 
from the original radial distance $r \lesssim 25$ au to its current orbit at 30 au. Here we show that the 
migration models with a very low orbital eccentricity of Neptune ($e_{\rm N} \lesssim 0.03$) do not explain KBOs with 
semimajor axes $50<a<60$ au, perihelion distances $q>35$ au and inclinations $i<10^\circ$. If $e_{\rm N} \lesssim 
0.03$ at all times, the Kozai cycles control the implantation process and the orbits with $q>35$ au end up having, 
due to the angular momentum's $z$-component conservation, $i>10^\circ$. Better results are obtained when Neptune's 
eccentricity is excited to $e_{\rm N} \simeq 0.1$ and subsequently damped by dynamical friction. The low-$e$ \& 
low-$i$ orbits at 50-60 au are produced in this model when KBOs are lifted from the scattered disk by secular 
cycles -- mainly the apsidal resonance $\nu_{8}$ -- near various mean motion resonances. These results give 
support to a (mild) dynamical instability that presumably excited the orbits of giant planets during Neptune's 
early migration.

\end{abstract}


\section{Introduction}

Previous studies of Kuiper belt formation envisioned dynamical models where Neptune maintained a very low orbital
eccentricity, comparable to the present $e_{\rm N} \simeq 0.01$, during its early migration (e.g., Malhotra 1993, 1995; 
Hahn \& Malhotra 2005; Gomes 2003). An instability model was subsequently suggested to explain the somewhat excited 
orbits of the outer planets (Tsiganis et al. 2005). In the original instability model, Neptune was scattered to 
a highly eccentric orbit ($e_{\rm N} \gtrsim 0.2$) that briefly overlapped with the Kuiper belt (e.g., Levison 
et al. 2008, Morbidelli et al. 2008, Gomes et al. 2018). Different arguments have been proposed in the past to rule 
out specific migration/instability regimes (e.g., Batygin et al. 2011, Wolff et al. 2012, Dawson \& Murray-Clay 
2012). For example, the {\it high}-eccentricity instability model with fast ($\sim 1$ Myr) subsequent circularization 
of Neptune's orbit does not reproduce the generally wide inclination distribution of KBOs, because there is not enough 
time to excite the orbital inclinations in this model (Nesvorn\'y 2015). Most modern studies of the instability 
therefore considered $e_{\rm N} \sim 0.1$ (e.g., Nesvorn\'y \& Morbidelli 2012; Kaib \& Sheppard 2016; Deienno et al. 
2017, 2018; Clement et al. 2020).

Here we highlight a new constraint that could be used to rule out the very-{\it low}-eccentricity migration of Neptune
($e_{\rm N} \lesssim 0.03$). 
The Outer Solar System Origins Survey (OSSOS; Bannister et al. 2018) identified a population of KBOs with semimajor axes 
$50<a<60$ au, perihelion distances $q>35$ au and inclinations $i<10^\circ$. We show that this population can easily be 
explained if Neptune reached eccentricity $e_{\rm N} \simeq 0.1$ during migration, perhaps due to its interaction with Uranus 
or a rogue planet (e.g., Nesvorn\'y 2011). In this model, bodies starting at $r \lesssim 30$ au are scattered by Neptune 
to orbits with $50<a<60$ au and $q \sim 30$ au, where they interact with various Neptune's mean motion 
resonances. The secular cycles, mainly associated with the $\nu_8$ resonance ($g = g_8$, where $g$ and $g_8$ are the 
apsidal precession frequencies of a KBO and Neptune, respectively), then act to decouple the scattered KBOs from Neptune 
and produce orbits with $50<a<60$ au and $q>35$ au. As the $\nu_8$ resonance leaves orbital inclinations unchanged, 
many of these bodies keep their initially low inclinations and end up with $i<10^\circ$, thus explaining the existence 
of high-$q$ and low-$i$ KBOs with $50<a<60$ au.

If, instead, $e_{\rm N} \lesssim 0.03$ during Neptune's migration, the $\nu_8$ resonance is not effective and scattered KBOs 
decouple from Neptune -- in absence of other external perturbations -- by the Kozai resonance (Kozai 1962; more 
specifically: by the Kozai resonance {\it inside} various mean motion resonances with migrating Neptune, see below).
As the quantity $\sqrt{1-e^2} \cos i$ is preserved during Kozai cycles, lowering the orbital eccentricity means 
that the inclination must increase. We find that this leads to a situation where bodies cannot reach the orbits 
with $50<a<60$ au, $q>35$ au and $i<10^\circ$, and use this result to argue against the very-low-eccentricity 
migration of Neptune. Alternatively, the KBOs with $50<a<60$ au, $q>35$ au and $i<10^\circ$ could have originated in 
a hypothetical, low-$e$ and low-$i$ disk at $50<r<60$ au that was subsequently disturbed by some process. We also 
discuss migration models with an elevated orbital inclination of Neptune, where the $\nu_{18}$ resonance ($s = s_8$, 
where $s$ and $s_8$ are the nodal precession frequencies of a KBO and Neptune, respectively) would affect 
orbital inclinations.

\section{Data, Model and Results}

\begin{figure}
\epsscale{0.8}
\plotone{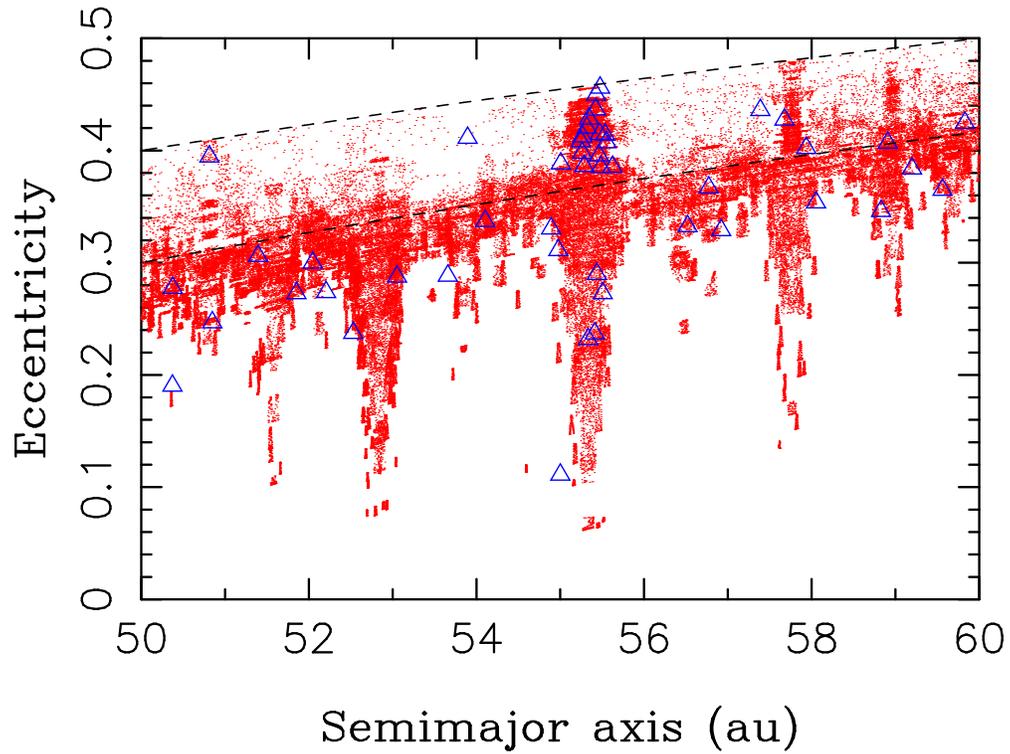}
\caption{The orbits of KBOs with $50<a<60$ au: the model orbits from s10/30j (Nesvorn\'y et al. 2002, 
$e_{\rm N}=0.1$; red dots) and OSSOS detections (blue triangles). The dashed lines show $q=30$ and 
35 au for a reference. The vertical columns of model objects correspond to the mean motion resonances 
with Neptune (e.g., 7:3 at 52.9 au, 5:2 at 55.4 au, 8:3 at 57.9 au).}
\label{case2a}
\end{figure}

The orbits of KBOs with $50<a<60$ au are shown in Fig. \ref{case2a}. OSSOS detected 52 KBOs in this region. There is
a cluster of objects with $q<35$ au in the 5:2 resonance ($a \simeq 55.4$ au; Malhotra et al. 2018). The 5:2 resonance 
was previously shown to host a large population of KBOs that can rival that of the 3:2 resonance (Gladman et al. 2012). This is not the 
subject of the present paper. Here we focus on the population of KBOs with $q>35$ au. There are 25 OSSOS 
bodies with $50<a<60$ au and $q>35$ au, of which 14 are classified by the OSSOS team as detached; the rest is classified 
as resonant (see Gladman et al. 2008 for a definition of different dynamical categories). The detached orbits are 
expected to be stable on long time intervals. We integrated all 25 OSSOS bodies with $50<a<60$ au and $q>35$ au and found that 
most of their orbits remain practically unchanged over 4.5 Gyr.\footnote{It is not clear whether the orbits are known 
well enough to be able to conclusively establish their long term stability. For example, the semimajor axis uncertainties 
are typically 0.002-0.04 au, which can matter if some of the orbits are very close to mean motion resonances. Five 
objects -- 2013 JH64, 2014 UG228, 2015 KT174, 2015 KV174, 2015 RU278 -- were scattered by Neptune and removed from 
the simulation. Of these, only 2014 UG228 has low orbital inclination ($i=3.5^\circ$).} This means that a great majority 
of these bodies must have evolved onto their current orbits during Neptune's early migration. For comparison, of the 27 
OSSOS bodies with $50<a<60$ au and $q<35$ au, 20 are resonant (mainly the 5:2 resonance), six are scattering and one as 
detached. This shows that $q=35$ au is a good division line, at least for $50<a<60$ au, between objects coupled to 
Neptune and those that are not.

\begin{figure}
\epsscale{0.45}
\plotone{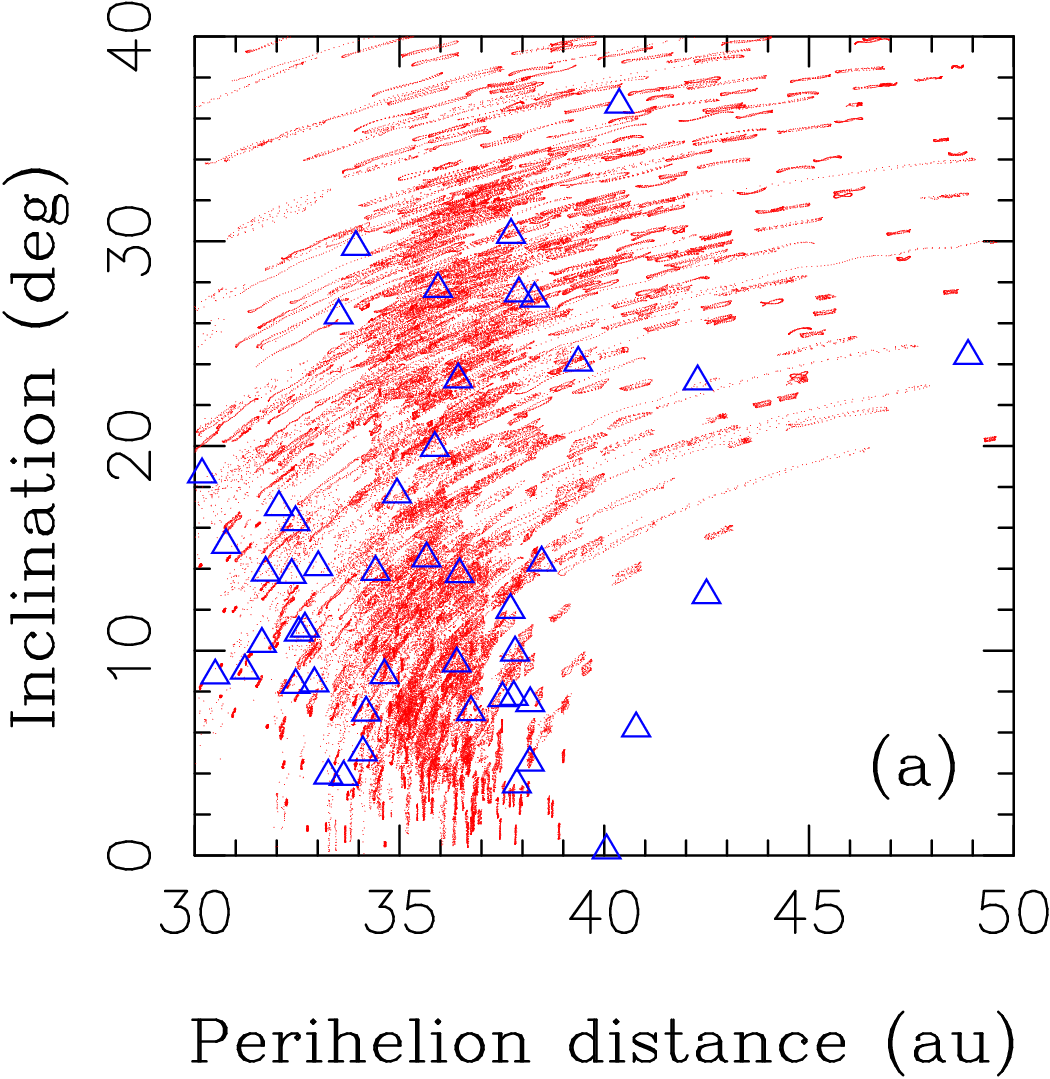}
\plotone{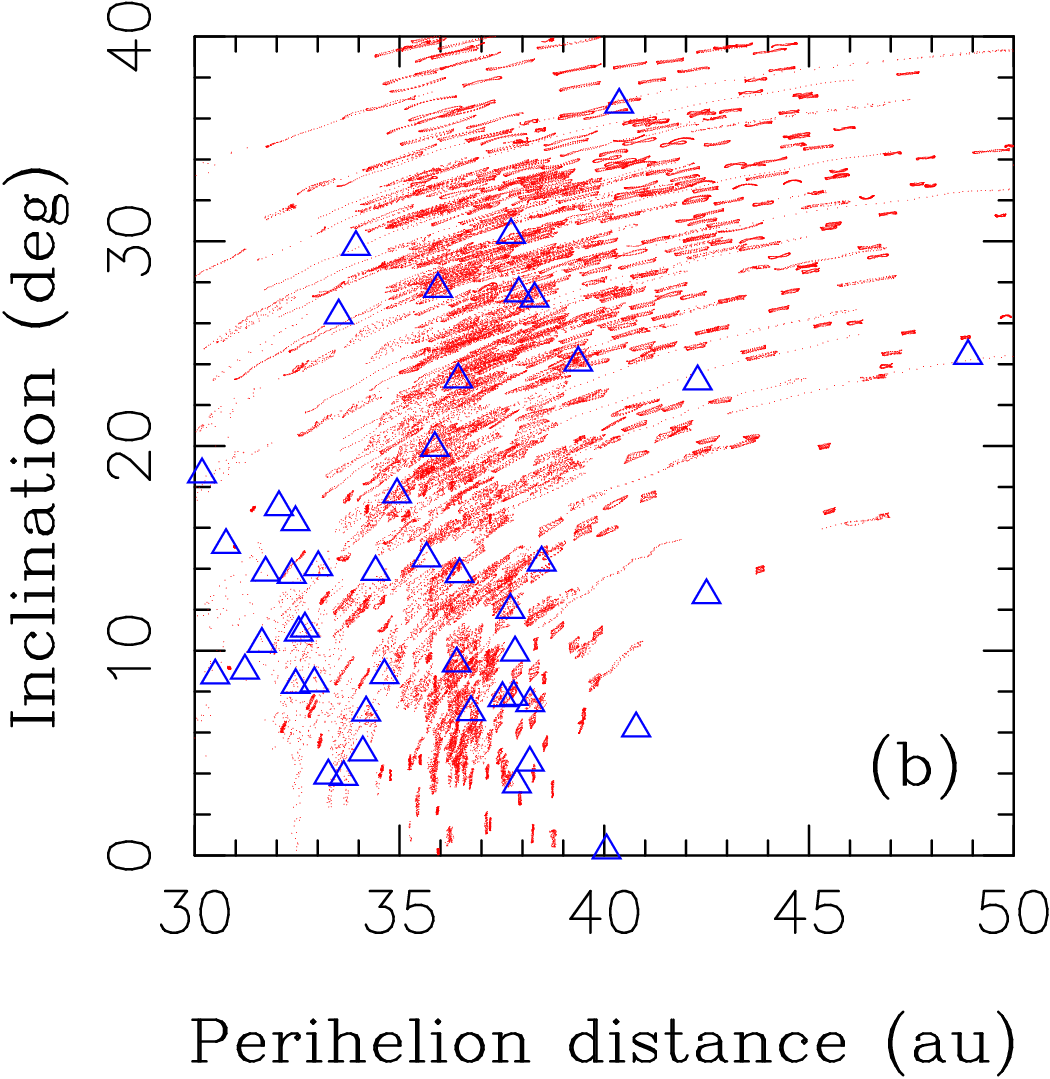}
\caption{The orbital inclinations and perihelion distances of KBOs with $50<a<60$ au: dynamical models 
s10/30j in panel a and s30/100j in panel b (red dots; $e_{\rm N}=0.1$, Nesvorn\'y et al. 2020). The model
orbits are shown in the time interval of 10 Myr around the present epoch. The blue triangles are KBOs 
detected by OSSOS (Bannister et al. 2018).}
\label{case}
\end{figure}

The inclination distribution of known KBOs with $50<a<60$ au and $q>35$ au is relatively wide with many orbits having 
$i<10^\circ$ (10 of 25, or 40\%; Fig. \ref{case}). Of these, five bodies are classified as resonant (two in 13:6, one in 
each 7:3, 8:3, 13:5, none in 5:2) and five as detached. The orbital eccentricities are too large for these bodies to 
accrete on their current orbits. They must have formed elsewhere and evolved onto current orbits early in the solar 
system history. Here we show how this constrains Neptune's migration.

The model results shown in Fig. \ref{case} were taken from Nesvorn\'y et al. (2020), where we conducted simulations of 
Neptune's early migration. Specifically, the results in Fig. \ref{case}a correspond to the s10/30j model (see Table 1 
in Nesvorn\'y et al. 2020) where Neptune migrated from $a_{\rm N,0}=24$ au to $a_{\rm N,0}=27.7$ au in the first 10 Myr, before 
its orbit was excited to $e_{\rm N}=0.1$ during the dynamical instability. Neptune's eccentricity was then slowly damped 
as Neptune migrated toward 30 au. The results in Fig. \ref{case}b correspond to the s30/100j model with a longer migration 
timescale and $e_{\rm N}=0.1$. In both models, Neptune's orbital inclination remained near its current value. We find 
that bodies are implanted onto orbits with $50<a<60$ au, $q>35$ au and $i<10^\circ$ by a combined effect of Neptune 
scattering, mean motion resonances and $\nu_8$. 

\begin{figure}
\epsscale{0.9}
\plotone{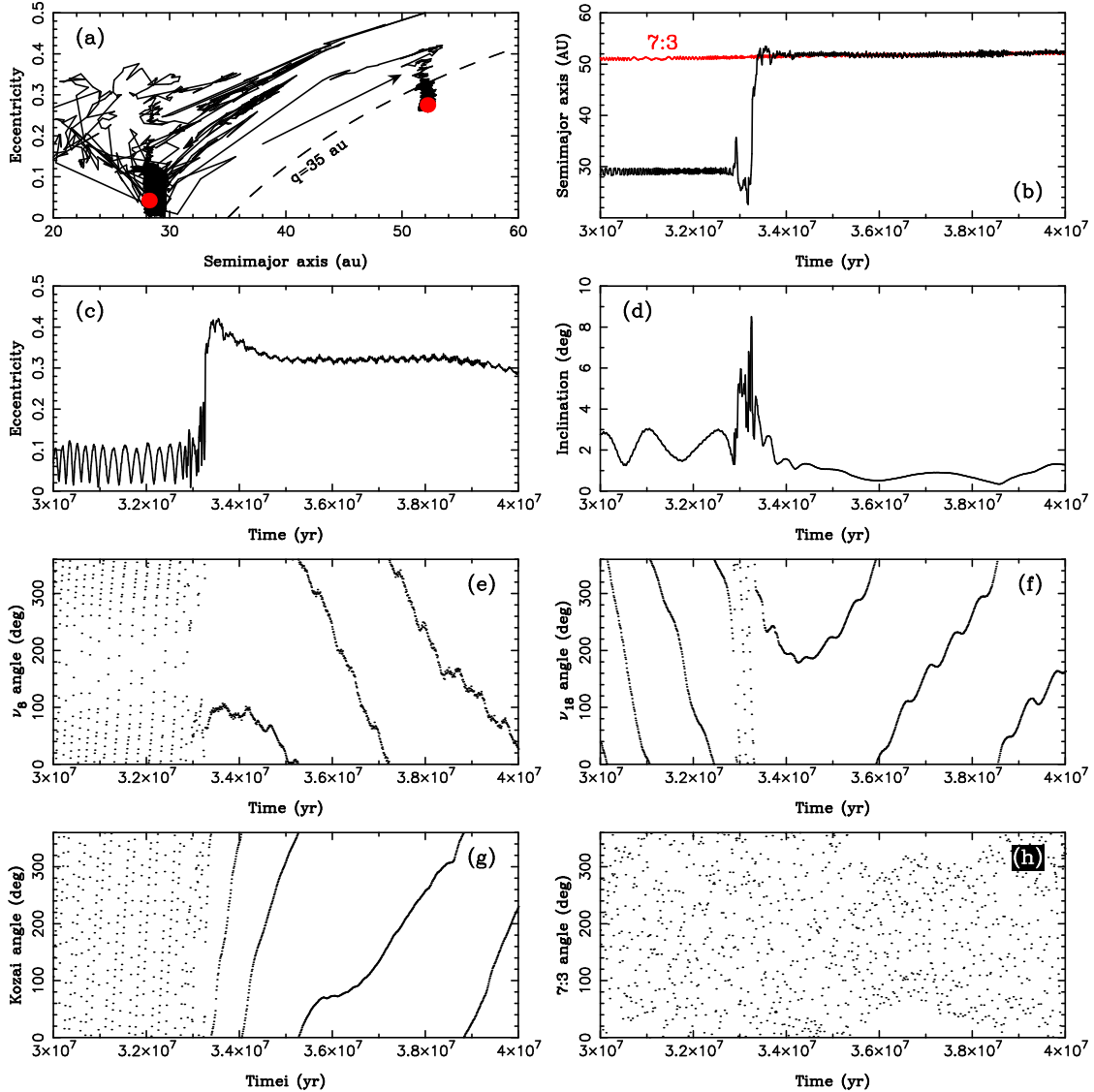}
\caption{The orbital history of a test body that started below 30 au and ended on a detached, low-inclination 
orbit with $a \simeq 52$ au (model s10/30j with $e_{\rm N}=0.1$ from Nesvorn\'y et al. 2020).
The red dots in panel (a) show the initial and final orbits. The red line in panel
(b) is the 7:3 resonance with Neptune. Various resonant angles are shown in panels (e)-(h): the $\nu_8$ resonance 
angle $\varpi-\varpi_{\rm N}$, where $\varpi$ and $\varpi_{\rm N}$ are the perihelion longitudes (panel e),
the $\nu_{18}$ resonance angle $\Omega-\Omega_{\rm N}$, where $\Omega$ and $\Omega_{\rm N}$ are the 
nodal longitudes (panel f), the Kozai resonance angle $\omega$, where $\omega$ is the perihelion argument
(panel g), and the 7:3 resonance angle $7 \lambda - 3 \lambda_{\rm N} - 4 \varpi$, where $\lambda$ and 
$\lambda_{\rm N}$ are the mean longitudes (panel h).}
\label{ex1}
\end{figure}

In the example shown in Fig. \ref{ex1}, taken from s10/30j with $e_{\rm N}=0.1$, the test body started below 30 au and 
was scattered by Neptune to $a \simeq 52$ au and $q \simeq 30$ au at $t \simeq 33.2$ Myr after the start of our 
simulation (panel a). It evolved onto an orbit with a very large libration amplitude in the 7:3 mean motion resonance 
with Neptune (panels b and h). The secular resonance $\nu_8$ (panel e) acted to decrease the orbital eccentricity 
(panels c), and the orbit was decoupled from Neptune. It was subsequently released from the 7:3 resonance and ended 
up on a stable orbit with $q \simeq 38$ au and $i \simeq 1^\circ$. We inspected 40 orbital histories of bodies that 
evolved onto the orbits with $50<a<60$ au, $q<35$ au and $i<10^\circ$ in the
s10/30j and s30/100j migration models and found that the $\nu_8$ resonance plays the dominant role in decoupling 
bodies from Neptune (most orbital evolutions are more complicated than the one shown in Fig. \ref{ex1}; different
mean motion resonances are involved). As the $\nu_8$ resonance does not affect orbital inclinations, orbits
can maintain their low inclinations as they evolve from the scattered orbits to $q>35$ au.  
 
We now compare the model distributions obtained in the s10/30j and s30/100j models with the OSSOS observations (Fig. 
\ref{case}). This is not an apple-to-apple comparison because the model represents an intrinsic orbital distribution, 
whereas the data are affected by observational biases (e.g., flux biases, pointing history, rate cuts and object leakage; 
Bannister et al. 2016). But that's not the point here. The point is that the migration model with $e_{\rm N} \simeq 0.1$ 
is capable of producing orbits with $50<a<60$ au, $q<35$ au and $i<10^\circ$ in the first place, and thus provides a 
straightforward explanation for this population. A detailed comparison with the OSSOS observations will require certain 
assumptions about the magnitude distribution of KBOs and the use of the OSSOS simulator (Lawler et al. 2018). This is 
left for future work. Here we just point out the orbits with low inclinations are preferentially picked up by the 
OSSOS observations near the ecliptic. 

\begin{figure}
\epsscale{0.45}
\plotone{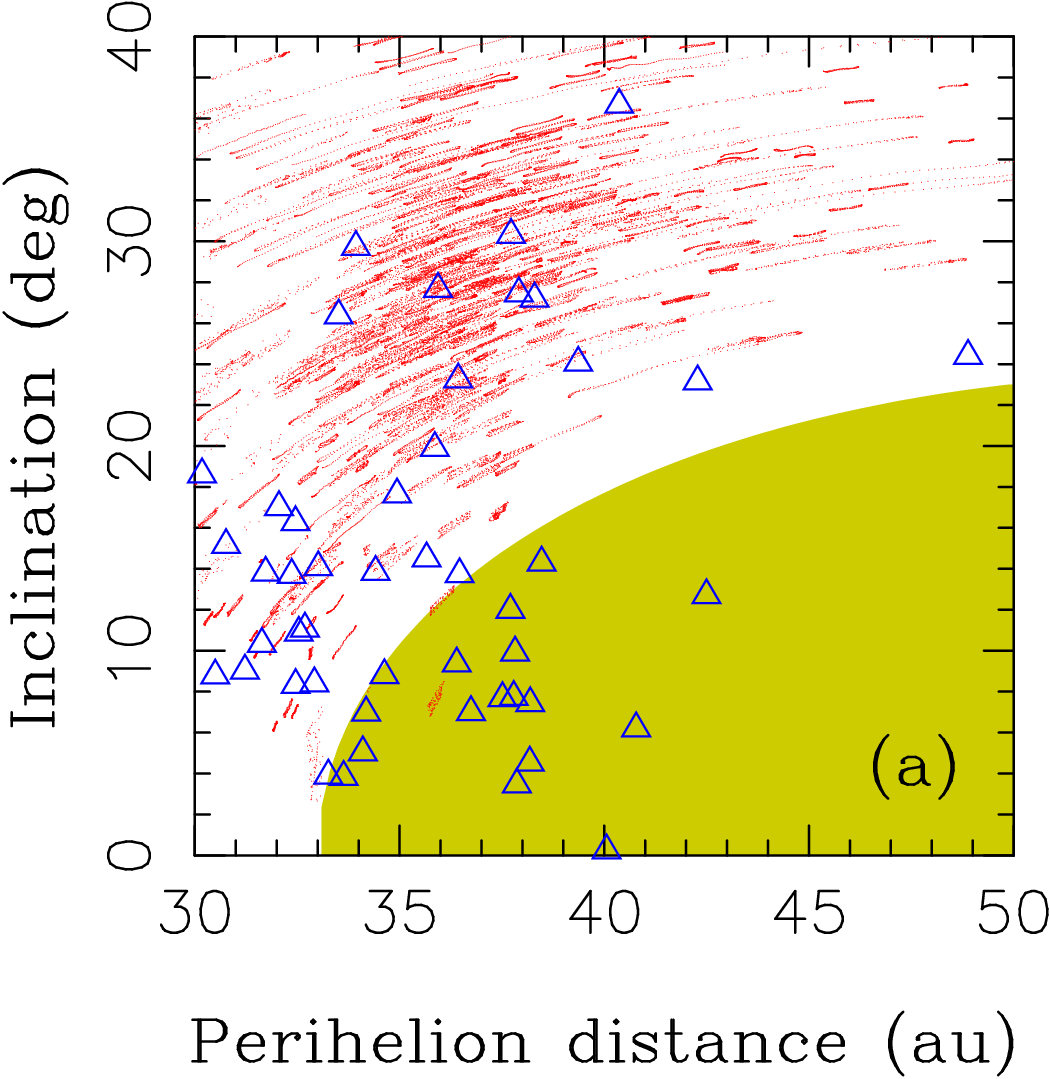}
\plotone{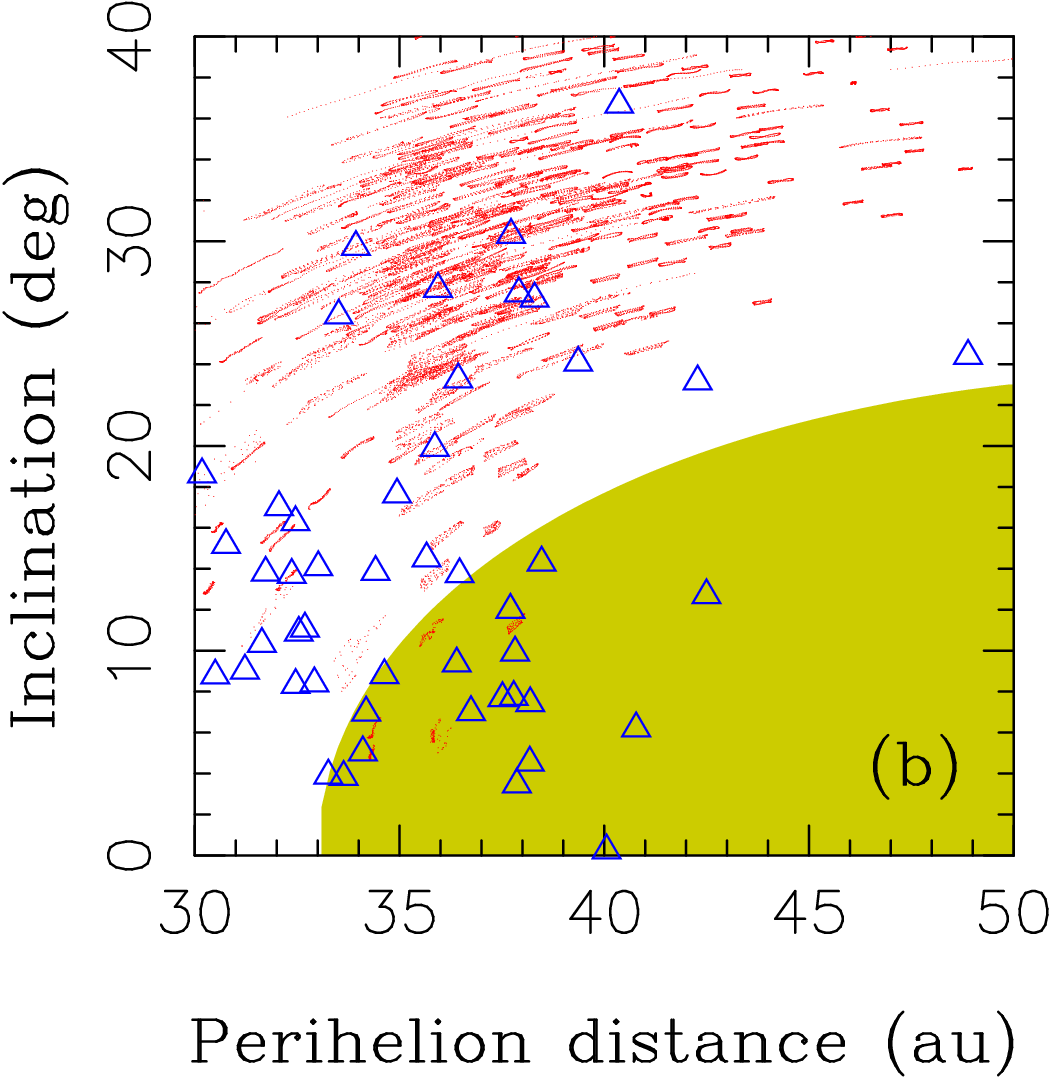}
\caption{The orbital inclinations and perihelion distances of KBOs with $50<a<60$ au: dynamical models 
s10/30j in panel a and s30/100j in panel b (red dots; $e_{\rm N} = 0$-0.02, Nesvorn\'y et al. 2020). The model
orbits are shown in the time interval of 10 Myr around the present epoch. The blue triangles are KBOs detected 
by OSSOS (Bannister et al. 2018). The shaded area (light green/dark yellow) is the excluded region that cannot 
be reached by orbits evolving from $q \sim 30$ au by the Kozai cycles.}
\label{zero}
\end{figure}

Figure \ref{zero} shows the results of the migration models where Neptune's eccentricity stayed low at all times ($e_{\rm N} 
\simeq 0$-0.02; no instability, otherwise identical to the s10/30j and s30/100j models discussed above). 
The model distributions are markedly different from the ones shown in Fig. \ref{case}. To understand this difference, 
note that the $\nu_8$ resonance becomes ineffective when Neptune's eccentricity is low (e.g., Morbidelli 2002, 
Sect. 11.2.3). This means that bodies cannot be decoupled from the scattered orbits by $\nu_8$. Instead, slower chaotic 
effects and the Kozai resonance come into action (Kozai 1962; see Sect. 11.2.2 in Morbidelli 2002 for the Kozai 
resonance {\it inside} mean motion resonances). The Kozai resonance was present in the models with $e_{\rm N}=0.1$ as 
well but there the decoupling mechanism was controlled by the stronger $\nu_8$ resonance.  

\begin{figure}
\epsscale{0.9}
\plotone{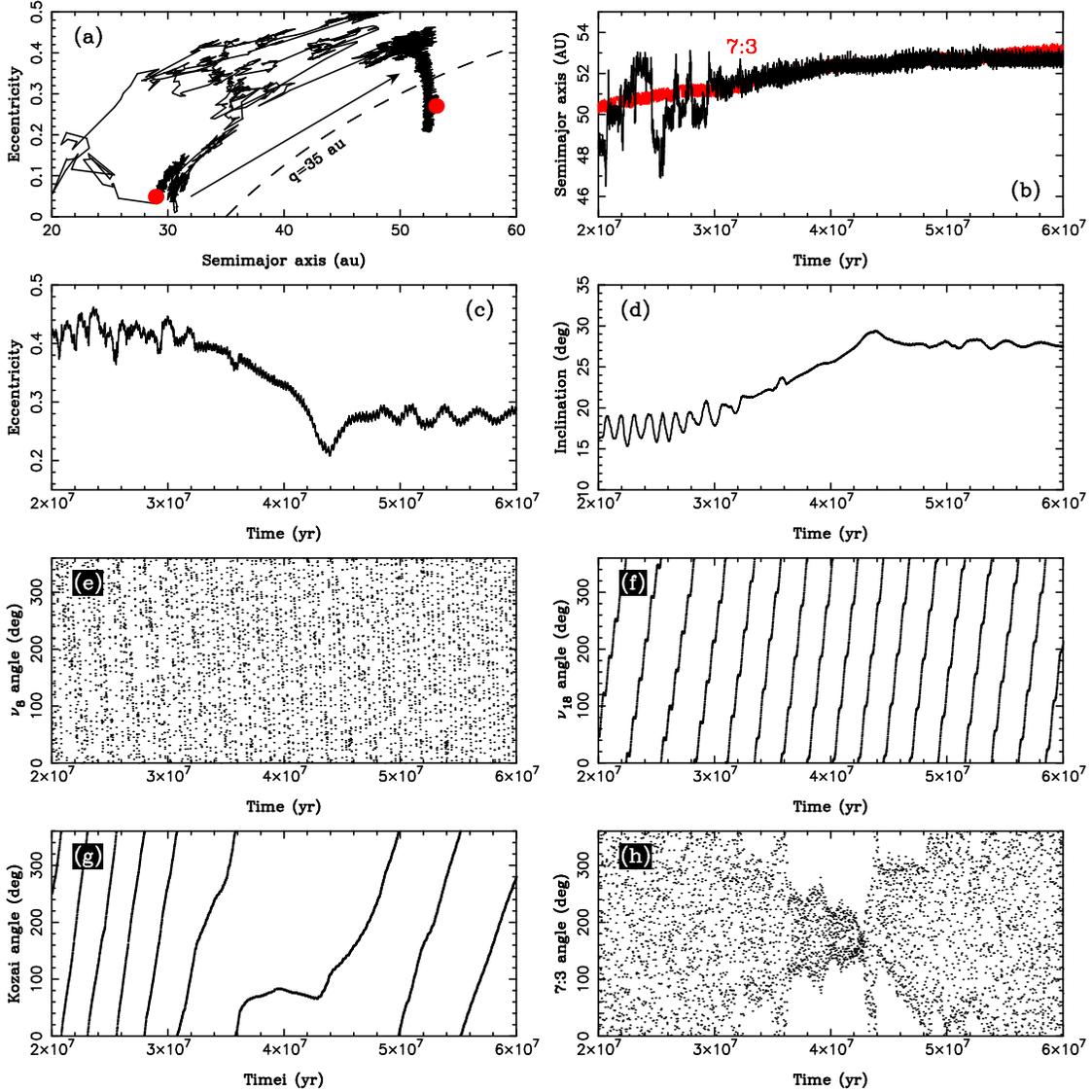}
\caption{The orbital history of a test body that started below 30 au and ended on a detached, high-inclination 
orbit with $a \simeq 52.5$ au (model s10/30j with $e_{\rm N} = 0$-0.02). See caption of Fig. \ref{ex1} for
a description of different panels. For the first $\sim30$ Myr the body is kicked around by encounters with 
the outer planets (panel a). It then evolves, near $t=36$ Myr, onto an orbit with a relatively small libration
amplitude in the 7:3 resonance (panels b and h). This triggers an incomplete Kozai cycle with $\omega$ 
nearly reversing its rotation just after $t=40$ Myr (panel g). As a result, the orbital eccentricity drops 
(panel c), the orbit becomes decoupled from Neptune, and the orbital inclination increases (panel d). 
The anti-correlated behavior of $e$ and $i$ is characteristic of the Kozai cycles (see main text). The orbit is 
eventually released from the migrating 7:3 resonance near $t=55$ Myr (panel b).}
\label{ex2}
\end{figure}

With $\nu_8$ gone for $e_{\rm N} \simeq 0$, the orbits affected by Kozai cycles conserve the $z$ component of the 
(scaled) angular momentum, $L_z=\sqrt{1-e^2} \cos i$. Thus, as the eccentricity is driven down during the decoupling 
process, the inclination must go up. An example of this evolution is shown in Fig. \ref{ex2}. This effectively 
defines an excluded region in the $(q,i)$ plane that cannot be reached by the decoupled orbits (even if they start 
on the scattered orbits with $i=0$).\footnote{In addition, in the absence of strong $\nu_8$-resonance effects, the
efficiency of the decoupling process drops. This then leads to a much smaller population of detached objects than
in the models with $e_{\rm N} \simeq 0.1$ (compare Figs. \ref{case} and \ref{zero}). The bodies that eventually evolve 
to $q>35$ au spend longer time on the Neptune-crossing orbits and tend to have larger inclinations. This further 
emphasizes the difference between Figs. \ref{case} and \ref{zero}.} Indeed, the model orbits do not evolve into this 
region if $e_{\rm N} \simeq 0$-0.02 (Fig. \ref{zero}). 

This is the heart of our argument: the observed orbits with $50<a<60$ au, $q>35$ au and $i<10^\circ$ cannot be explained 
in the migration model with $e_{\rm N} \simeq 0$-0.02. Their existence implies that the $\nu_8$ resonance was effective 
during Neptune's migration, rules out the migration model with $e_{\rm N} \simeq 0$-0.02, and gives support to the models 
with (mild) dynamical instabilities ($e_{\rm N} \simeq 0.1$; e.g., Nesvorn\'y 2011, Nesvorn\'y \& Morbidelli 
2012).\footnote{See Nesvorn\'y et al. (2020) for other KBO populations obtained in the model with $e_{\rm N}=0.1$.} 
To get better hold of the critical eccentricity of Neptune implied by this argument, we performed two additional 
simulations: (i) s10/30j with $e_{\rm N}=0.03$, and (ii) another one with shorter migration timecales ($\tau=5$ and 
10 Myr, see Nesvorn\'y et al. 2020 for a definition of these parameters) and $e_{\rm N}=0.1$. The results of (i) are 
similar to the case with $e_{\rm N}=0$-0.02, which allows us to estimate the critical eccentricity 
$e_{\rm N} \sim 0.05$. From (ii), we speculate that the critical eccentricity may show a weak dependence on the 
timescale of Neptune's migration with shorter timescales implying slightly larger thresholds.
  
\section{Discussion}

Arguments similar to those described above can be made based on the orbits of detached orbits with $a>60$ au.
We are less confident in drawing inferences from that population, because: (1) the population of 
detached bodies with $a>60$ au is still relatively small, and (2) it may have been affected by stellar 
encounters, planet 9, or other external perturbations not included in our model. We thus leave a careful study
of that population for future work. As for $a<50$ au, the observed KBOs are thought to be a mixture of dynamically 
`cold' ($i<5^\circ$) and `hot' populations ($i>5^\circ$), respectively representing planetesimals that formed 
at $r \simeq 42$-47 au (i.e., between the 3:2 and 2:1 resonances with Neptune) and $r<30$ au (i.e., below
the current orbit of Neptune; see Morbidelli \& Nesvorn\'y 2020 for a review). It is difficult to build a case 
on this population, because most of the low-$i$ bodies found there did not evolve through the scattered disk. 
Instead, they were born and survived with $40<a<47$ au and $i<5^\circ$. 

This raises a question whether the observed population of KBOs with $50<a<60$ au, $q>35$ au and $i<10^\circ$ 
could simply be a continuation of the cold disk from below the 2:1 resonance to $r>50$ au. The main argument
against this possibility is that the cold KBOs at $r \simeq 42$-47 au remained with $i<5^\circ$. If the same
population existed beyond 50 au, it would preferentially be seen by OSSOS (which is strongly biased, due to 
its observing strategy, toward discovering objects with $i \sim 0$). The orbits with $50<a<60$ au instead have
a broad inclination distribution and do not show any hint of bimodality seen inward of the 2:1 resonance.
Perhaps, then, some unknown dynamical mechanism (3:1 resonance?, rogue planets?) have strongly excited the orbits 
starting with $50<a<60$ au and $e \sim i \sim 0$. In that case, however, it would have to be demonstrated that 
the mechanism can excite orbital inclinations for $a>50$ au but leave $i<5^\circ$ for $40<a<47$ au. 
   
The detached orbits can potentially be generated from the scattering disk by rogue planets scattered off of 
Neptune (Gladman \& Chan 2006). For this to happen, the largest rogue planet would need to have mass $\gtrsim 1$ 
$M_{\rm Earth}$, where $M_{\rm Earth}$ is the Earth mass (Gladman \& Chan 2006). Shannon \& Dawson (2018) placed 
an upper limit on the number of Earth-mass planets in the original outer disk. From the preservation of ultra
wide KBO binaries they found that $<3$ Earth-mass planets could have been initially present (68\% statistical 
probability), and from the low orbital inclinations of cold classicals they found $<1$ Earth-mass planets (68\% 
probability). Only a fraction of these planets would be scattered outward by Neptune and could contribute to
the construction of the detached disk. Unfortunately, detailed predictions of the rogue-planet model for the 
orbital distribution of KBOs at 50-60 au are not available at this moment. 
 
Additional possibility is related to the effects of the $\nu_{18}$ resonance. Volk \& Malhotra (2019) showed 
that $\nu_{\rm 18}$ can increase the orbital inclinations of bodies implanted in the 3:2 resonance, if Neptune's
inclination was somewhat elevated above its current value (proper $i_{\rm N} \simeq 0.7^\circ$). If that's the case,
assuming $e_{\rm N} \sim 0$ and large $i_{\rm N}$, we speculate that the $\nu_{18}$ resonance (near and inside 
various mean motion resonances; see Nesvorn\'y \& Roig 2000, 2001; Volk \& Malhotra 2019) may have scrambled the 
inclination distribution of bodies with $q>35$ au. This could potentially change the inclination distribution 
and explain the observed KBO population with $50<a<60$ au, $q>35$ au and $i<10^\circ$. It is unknown, however, 
what could cause high $i_{\rm N}$ and leave low $e_{\rm N}$ during Neptune's migration, because mean motion 
resonances and/or scattering tend to excite $e_{\rm N}$. 

Finally, as an important caveat, we emphasize that our exploration of parameter space is incomplete. 
In the simulations presented here, we varied Neptune's eccentricity ($e_{\rm N}=0$, 0.03 and 0.1) and the migration/damping  
timescales ($\tau=5$-10, 10-30 and 30-100 Myr). We did not test the dependence on other parameters, such as $i_{\rm N}$ 
(see the discussion above), the relative position of planets during migration, etc. For example, the location 
of the $\nu_8$ resonance can be sensitive, in general, to the detailed configuration of planets. The $\nu_8$ 
resonance, however, always occurs near and inside of mean motion resonances with Neptune where the apsidal 
precession rate of orbits has a singularity (Kne\v{z}evi\'c et al. 1991). Thus, given that the implantation
mechanism identified here relies on the proximity to a mean motion resonance, the detailed behavior of planets 
should be inconsequential for the main argument presented in this work. 

\section{Conclusions}

We find that the population of KBOs with $50<a<60$ au, $q>35$ au and $i<10^\circ$ can be explained if Neptune's 
orbit was excited to $e_{\rm N} \simeq 0.1$ during planetary migration and subsequently dumped to the current 
$e_{\rm N} = 0.01$ by dynamical friction. If, instead, Neptune maintains a very low orbital eccentricity, 
$e_{\rm N} \lesssim 0.03$, at all times, these low-$i$ KBOs are not obtained in the simulations. Our
results can be interpreted to give support for a dynamical instability in the early outer solar system. 
Questions remain, however, related to the cause and orbital behavior of planets during the instability. 

The gravitational interaction of Uranus and Neptune, mainly when these planets cross a mutual mean motion 
resonance during their migration, can excite Neptune's eccentricity but it is uncommon for this process 
to generate $e_{\rm N} > 0.05$. The planetary encounters between ice giants, with Neptune being scattered off 
of a rogue planet, lead to $e_{\rm N} \sim 0.1$ quite often (Nesvorn\'y 2011, Nesvorn\'y \& Morbidelli 2012, 
Batygin et al. 2012). The planetary encounters of Neptune with Jupiter or Saturn typically lead to $e_{\rm N} 
\gtrsim 0.2$ (Tsiganis et al. 2005, Levison et al. 2008), but this model struggles to match other constraints 
(e.g., Nesvorn\'y \& Morbidelli 2012, Nesvorn\'y 2015). From these considerations, we favor the mild 
instability model with a large rogue planet.

\acknowledgements

The work of D.N. was supported by the NASA Emerging Worlds program. We thank the anonymous reviewer for helpful 
comments on the submitted manuscript.

\clearpage

\end{document}